# How can we kill cancer cells: insights from the computational models of apoptosis

Subhadip Raychaudhuri
Department of Biomedical Engineering, Biophysics Graduate Group, Graduate Group in Immunology, and Graduate Group in Applied Mathematics, University of California, Davis, California 95616
Address Correspondence: raychaudhuri@ucdavis.edu

**Abstract**

Cancer cells are widely known to be protected from apoptosis, which is a major hurdle to successful anti-cancer therapy. Over-expression of several anti-apoptotic proteins, or mutations in pro-apoptotic factors, has been recognized to confer such resistance. Development of new experimental strategies, such as *in silico* modeling of biological pathways, can increase our understanding of how abnormal regulation of apoptotic pathway in cancer cells can lead to tumour chemoresistance. Monte Carlo simulations are in particular well suited to study inherent variability, such as spatial heterogeneity and cell-to-cell variations in signaling reactions. Using this approach, often in combination with experimental validation of the computational model, we observed that large cell-to-cell variability could explain the kinetics of apoptosis, which depends on the type of pathway and the strength of stress stimuli. Most importantly, Monte Carlo simulations of apoptotic signaling provides unexpected insights into the mechanisms of fractional cell killing induced by apoptosis-inducing agents, showing that not only variation in protein levels, but also inherent stochastic variability in signaling reactions, can lead to survival of a fraction of treated cancer cells.

**Introduction**

Induction of apoptosis by chemotherapeutics is considered as one of the major anti-cancer effects leading to inhibition of tumour growth. Over the past years our understanding of signaling pathways associated with induction of apoptosis, and knowledge on executioners of apoptosis, has substantially increased. Recently, cell-to-cell stochastic variability has become central to apoptotic cell death signaling (1-6). Computational models are well suited to provide mechanistic insight into the system level regulations of apoptosis signaling and its large cellular variability. Studies that are possible using *in silico* approaches might be inaccessible by other techniques. Recent novel findings in the area of apoptotic cell death signaling can have far-reaching implications in cancer biology and therapy. Our computational efforts, in synergy with parallel biological experiments, attempt to explore some of the fundamental issues in cancer biology within this new paradigm of apoptosis signaling.

**Developing computational models of cell death signaling for normal and cancer cells**

Monte Carlo models are generally suitable for simulating inherent stochasticity of signaling reactions in complex signaling pathways (1,7). Our recent work has elucidated that cell-to-cell stochastic variability is a fundamental characteristic of apoptosis signaling and a significant part of such variability can arise due to inherent stochastic fluctuations of chemical reactions (1,6,8,9). Even when the intrinsic stochastic variability is not dominant, Monte Carlo models have the advantage over ordinary differential equation (ODE) based models, as they can simulate spatial heterogeneity in an explicit manner. Examples of such spatial localizations in apoptosis signaling include translocation of activated Bax molecules onto mitochondrial outer membrane, release of cytochrome *c* from mitochondria to cytosol and redistribution of Apaf-1 in the cytosol, all of which depend on the cell type and the level of Bcl-2 proteins (10,11). In addition, we could easily incorporate realistic variations in (i) protein concentrations that may arise from stochastic gene regulations (3,12-15), and (ii) reaction rate constants, for example, due to variations in pH in the cytosolic environment. In our initial studies we grouped functionally redundant proteins so that a single representative protein simulates all proteins of similar function that are possibly expressed within a given cell type. For example, apoptotic inhibitor Bcl-2 captured the effect of all the inhibitory Bcl-2 like proteins (1,9). In the future we plan to simulate a more cell-type-specific signaling network of apoptosis, results of which can be readily compared with data obtained from parallel biological experiments for specific cell types. Such an expanded signaling network will often involve signaling species with low concentrations at initial time, or dynamically generate a few molecules due to specific inhibitory reactions, leading to inherent stochastic fluctuations that can be best studied using stochastic approaches.

**Differential signaling through extrinsic and intrinsic pathways of apoptosis**

Apoptosis is regulated through two distinct signaling pathways that are joined in a global loop structure as both pathways converge on activation of effecter Caspase-3 (10). The extrinsic (also called the type 1) pathway directly activates Caspase-3 by enzymatic reactions catalyzed by activated Caspase-8 molecules. The intrinsic (also called the type 2) pathway is regulated by mitochondrial cytochrome *c* release and apoptosome formation. We can assume that three local signaling modules coordinate apoptosis in the type2 pathway (Figure 1). Cell death can be induced by an apoptotic stimulus acting at any of the three different signaling modules (or right before them) of the apoptotic pathway (Figure 1): (i) death ligand binding and Caspase-8 activation, (ii) Bax activation by BH3 only proteins, (iii) $Ca^{2+}$ release that activates only the post-mitochondrial events. Our initial studies showed that, for the case of Caspase-8 mediated apoptosis, concentration of active Caspase-8 decides between the two pathways of apoptosis. Membrane reorganization, such as clustering of death receptors in raft signaling domains, determines the level of Caspase-8 activation in a cell type specific manner (16). For large concentrations of Caspase-8, direct Caspase-3 activation occurs in a fast (~ minutes) deterministic manner. Such rapid activation of apoptosis has been observed in various cell types due to Fas ligand binding to Fas receptor (17). Decrease in the strength of an apoptotic stimulus begins to activate the intrinsic pathway (Figure 2), as the rate constant for Caspase-8-Bid interaction is higher than that for Caspase-8-capsase 3 association (1,18).

**Large cell-to-cell variability through the intrinsic (mitochondrial) pathway can explain slow apoptosis**

We observe slow apoptosis (~ hours) when low concentrations of Caspase-8 are used in our simulations. We also find large cell-to-cell stochastic variability in the case of slow apoptosis. Similar signaling behavior is observed, irrespective of Caspase-8 concentration used, when we set the kinetic constants for the type 1 pathway to zero, confirming that slow apoptosis is a characteristic of the intrinsic pathway of apoptosis signaling. Such a study of pure type 2 apoptosis can be carried out *in silico* in a clean manner but will be difficult to achieve in biological experiments. When we perturbed the intrinsic apoptotic pathway downstream of Caspase-8 activation, we kept observing slow apoptosis with large cell-to-cell variability. Our results seem to explain very slow (~ 1 – 100 hours) apoptosis observed experimentally under a variety of conditions, for various types of cells and apoptotic stimuli, including under oxidative stress (2-6,9,10,19,20). Additional variations in protein concentrations in our simulations act in tandem with intrinsic stochasticity of signaling reactions to enhance cell-to-cell variability in apoptosis. Caspase-3 activation occurs in an all-or-none (digital) manner for single cells implicating signaling amplification of a weak stimulus through the intrinsic pathway. However, the information of strength of the stimulus is contained in the time-to-death and its cell-to-cell variability. Large cell-to-cell variability with all-or-none type Caspase-3 activation, as observed in our simulations, resulted in bimodal probability distributions for Caspase-3 activation that are thought to be characteristic of apoptosis signaling through the intrinsic pathway (1). Later experiments confirmed existence of such bimodal probability distributions in Caspase-3 activation (2,3,6,9).

**Minimal model of a signaling network demonstrates cell-to-cell variability in apoptosis in a cell-type independent manner**

In parallel, we derived a minimal model of a signaling network that is designed to sense an external stimulus and respond to it in an adaptive manner (8). This minimal network is derived based on some simple assumptions on its signaling response without any prior knowledge of the apoptotic pathway. A three-step fast-slow-fast pathway in the minimal network was shown to be sufficient to generate large cell-to-cell variability as observed in our Monte Carlo simulations of the intrinsic pathway of apoptosis (1). This minimal network also captures the change from rapid deterministic to slow stochastic signaling as the strength of the stimuli is varied, and a quantitative estimation of the threshold stimulus is obtained. This could be potentially important if one wants to engineer cancer cells to convert from type 2 to type 1 for fast apoptotic activation. The crucial slow reaction in the intermediate step of the minimal network can mimic the slow activation of Bax or the apoptosome formation in apoptosis signaling. Thus we can infer that some of the pertinent qualitative features of apoptosis signaling, as observed in our Monte Carlo simulations, are cell type independent. Such a conclusion is significant given the fact that cellular protein levels and even the type of molecules present in the apoptotic pathways vary significantly among cell types (3,10,12). Cancer cells are known to over-express a variety of apoptotic inhibitors, which confer them unusual resistance to apoptosis (21-24). The level of over-expression varies significantly among cancer sub-types and even among patients having the similar sub-types (25,26).

**How apoptotic inhibitors provide protection to cancer cells: implications for cancer therapy**

In a recent study we have shown that over-expression of Bcl-2 like proteins can slow down apoptosis and increase cell-to-cell stochastic variability (9). A high Bcl-2 level allows activation of only a few Bax molecules under apoptotic stimuli and thus dynamically generates mechanisms for stochastic fluctuations caused by small number of molecules. Interestingly, cancer cells are often primed for death by increasing the levels of apoptotic BH3 proteins but apoptosis in such cells is kept in check by continuous inhibition by anti-apoptotic Bcl-2 like proteins (22-24). Bcl-2 binds with several pro-apoptotic molecules creating a local loop structure (signaling module 2) in the intrinsic pathway that leads to non-linear and stochastic effects in its inhibitory action. Our simulations demonstrate that, beyond a threshold level, Bcl-2 imparts a strong inhibitory effect on apoptosis and thus can explain apoptosis resistance of cancer cells. For normal cells, having over-expressed Bcl-2 proteins, prolonged time-to-death might provide an opportunity for a particularly slow cell to acquire tumor initiating features. Behavior similar to tBid-Bcl-2-Bax signaling (in module 2) might be observed downstream of mitochondria (in signaling module 3) where higher Apaf-1 level might make cancer cells prone to apoptotic death, but such an effect can be negated by dominant effect of increased XIAP levels (27). XIAP also has multiple binding partners in a local loop network structure and contributes to generation of highly non-linear and stochastic signaling. Hence pre- and post- mitochondrial events in the intrinsic pathway are heavily regulated by two different loop network structures in two distinct signaling modules (Figure 1). Computational models are well suited to elucidate mechanisms of non-linear and stochastic signaling through those signaling modules. As a result, such models can help design optimal strategies to perturb those signaling modules by making use of the inherent apoptotic vulnerability of cancer cells. Initial simulations show increased cell death only for cancer cells having over-expressed BH3 protein Bid (unpublished observations), under a single agent treatment scenario such as under the action of Bcl-2 inhibitor HA14-1 (28-30). Our computational studies can clarify the basis of such inherent vulnerability of cancer cells for all three signaling modules (Figure 1). However, targeting only a single module (Figure 1), for example ligation of death receptors at the signaling module 1, will provide an opportunity for a significant number of cells to escape death. Such fractional killing of cancer cells occurs not only due to cellular variations in protein levels but also from inherent stochastic variability in signaling reactions (9). Computational modeling was well suited to establish that inherent stochastic variability by itself, even when all the other cellular parameters remain identical, can generate large cell-to-cell variability comparable to that observed in apoptosis activation experiments (9). Such large cell-to-cell variability in time-to-death provides an opportunity for opposing growth signals to up-regulate downstream apoptotic inhibitors such as XIAP. This is particularly relevant as apoptosis activation under chemotherapeutic treatment can be slow enough to allow synthesis of inhibitor proteins of varying concentrations through stochastic gene regulations. Targeting multiple signaling modules simultaneously using combinatorial treatments can be effective in reducing stochastic effects and fractional killing of cancer cells. Computational studies can provide us with a range of concentrations for optimal induction of apoptosis in a combinatorial treatment scenario and can guide the design of biological experiments. We are currently exploring the combined effect of HA14-1, an inhibitor of Bcl-2 (28-30), and embelin (27), an inhibitor of XIAP, that can induce apoptotic collapse in cancer cells.

**Acknowledgements**
We thank S.C. Das and Dr. J. Skommer for help with preparing this review article. We also thank Dr. J. Skommer and Dr. T. Brittain for helpful discussions.

**Figures**

Figure 1: Schematics of the apoptosis signaling pathway that indicates existence of three distinct signaling modules in the intrinsic pathway of apoptosis. Apoptosis can be activated at various locations in the intrinsic pathway. We also show some of the targets of cancer drugs in that pathway.

Figure 2: Time course of caspase-3 activation for low concentration of caspase-8 (< 5 nanomolar). Data is shown for five individual cells. Arrow indicates switch to the intrinsic pathway of apoptosis at the level of single cells.

**References**
1. Raychaudhuri, S., Willgohs, E., Nguyen, T.N., Khan, E.M., Goldkorn, T. (2008) Monte Carlo simulation of cell death signaling predicts large cell-to-cell stochastic fluctuations through the type 2 pathway of apoptosis. Biophys. J. *95*, 3559-62.
2. Albeck, J.G., Burke, J.M., Spencer, S.L., Lauffenburger, D.A., Sorger, P.K. (2008) Modeling a snap-action, variable-delay switch controlling extrinsic cell death. PLoS Biol. *6*, 2831-52.
3. Spencer, S.L., Gaudet, S., Albeck, J.G., Burke, J.M., Sorger, P.K. (2009) Non-genetic origins of cell-to-cell variability in TRAIL-induced apoptosis. Nature *459*, 428-32.
4. Düssmann, H., Rehm, M., Concannon, C.G., Anguissola, S., Würstle, M., Kacmar, S., Völler, P., Huber, H.J., Prehn, J.H. (2010) Single-cell quantification of Bax activation and mathematical modelling suggest pore formation on minimal mitochondrial Bax accumulation. Cell Death Differ. *17*, 278-90.
5. Luo, K.Q., Yu, V.C., Pu, Y., Chang, D.C. (2003) Measuring dynamics of caspase-8 activation in a single living HeLa cell during TNFalpha-induced apoptosis. Biochem Biophys Res. Commun. 304, 217-222.
6. Raychaudhuri, S., Skommer, J., Henty, K., Birch, N., Brittain, T. (2010) Neuroglobin protects nerve cells from apoptosis by inhibiting the intrinsic pathway of cell death. Apoptosis, 15, 401-411.
7. Raychaudhuri, S., Tsourkas, P.K., Willgohs, E. in Biophysics Fundamentals, edited by Jue T. (2009 Humana, vol 1, chap. 3).
8. Raychaudhuri, S. (2010) Minimal model of a signaling network elucidates cell-to-cell stochastic variability in apoptosis. *PLoS One* (In press).
9. Skommer, J., Brittain, T., Raychaudhuri, S. Bcl-2 inhibits apoptosis by increasing the time-to-death and intrinsic cell-to-cell variations in the mitochondrial pathway of cell death. *Apoptosis* (In Press).
10. Signaling Pathways in Apoptosis, edited by Watters D. and Lavin M. (1999) hardwood academic publishers.
11. Ruiz-Vela, A., Albar, J.P., and Martinez-A, C. (2001) Apaf-1 localization is modulated indirectly by Bcl-2 expression. FEBS Letters 510, 79-83.
12. Sigal, A., Milo, R., Cohen, A., Geva-Zatorsky, N., Klein, Y., Liron, Y., Rosenfeld, N., Danon, T., Perzov, N., Alon, U. (2006) Variability and memory of protein levels in human cells. Nature *444*, 643-6.
13. McAdams, H.H., Arkin, A. (1999) It's a noisy business! Genetic regulation at the nanomolar scale. Trends Genet. *15*, 65-69.
14. Elowitz, M.B., Levine, A.J., Siggia, E.D., Swain, P.S. (2002) Stochastic gene expression in a single cell. Science *297*, 1183-6.
15. Fedoroff, N., Fontana, W. (2002) Genetic networks: small numbers of big molecules. Science *297*, 119-1131.
16. Scott, F.L., Stec, B., Pop, C., Dobaczewska, M.K., Lee, J.J., Monosov, E., Robinson, H., Salvesen, G.S., Schwarzenbacher, R., Riedl, S.J. (2009) The Fas/FADD death domain complex structure unravels signaling by receptor clustering. Nature 457, 1019-1022
17. Scaffidi, C., Fulda, S., Srinivasan, A., Friesen, C., Li, F., Tomaselli, K.J., Debatin, K.M., Krammer, P.H., Peter M.E.. (1998) Two CD95 (APO-1/Fas) signaling pathways. EMBO J. 17, 1675-87.
18. Hua, F., Cornejo, M.G., Cardone, M.H., Stokes, C.L., Lauffenburger, D.A. (2005) Effects of Bcl-2 levels on Fas signalling-induced caspase-3 activation: molecular genetic tests of computational model predictions. J. Immunol. 175:985-995.

19. Goldkorn, T., Ravid, T., Khan, E.M. (2005) Life and death decisions: ceramide generation and EGF receptor trafficking are modulated by oxidative stress. Antioxidant and Redox Signaling 7, 119-128.
20. Spierings, D., McStay, G., Saleh, M., Bender, C., Chipuk, J., Maurer, U., Green, D.R. (2005) Connected to death: the (unexpurgated) mitochondrial pathway of apoptosis. Science 310, 66-67.
21. The Biology of Cancer, by Weinberg R.A. (2007) Garland Science, Taylor & Francis Group.
22. Letai, A., Sorcinelli, M.D., Beard, C., Korsmeyer, S.J. (2004) Antiapoptotic BCL-2 is required for maintenance of a model leukemia. Cancer Cell *6*, 241-9.
23. Certo, M., Del Gaizo Moore, V., Nishino, M., Wei, G., Korsmeyer, S., Armstrong, S.A., Letai, A. (2006) Mitochondria primed by death signals determine cellular addiction to antiapoptotic BCL-2 family members. Cancer Cell *9*, 351-65.
24. Skommer, J., Wlodkowic, D., Deptala, A. (2007) Larger than life: Mitochondria and the Bcl-2 family. Leuk. Res. *31*, 277-86.
25. Bradbury, D.A., Russel, N.H. (1995) Comparative quantitative expression of bcl-2 by normal and leukaemic myeloid cells. B. J. Haemot. 9, 374-379.
26. Porwit-MacDonald, A., Ivory K., Wilkinson, S., Wheatley, K., Wong, L., Janossy, G. (1995) Bcl-2 protein expression in normal human bone marrow precursors and in acute myelogenous leukemia. Leukemia 9, 1191-1198.
27. Nikolovska-Coleska, Z., Xu, L., Hu, Z., Tomita, Y., Li, P., Roller, P.P., Wang, R., Fang, X., Guo, R., Zhang, M., Lippman, M.E., Yang, D., Wang, S. (2004) Discovery of embelin as a cell-permeable, small-molecular weight inhibitor of XIAP through structure-based computational screening of a traditional herbal medicine three-dimensional structure database. J Med Chem. 47, 2430-40.
28. Skommer, J., Wlodkowic, D., Mättö, M., Eray, M., Pelkonen, J. (2006) HA14-1, a small molecule Bcl-2 antagonist, induces apoptosis and modulates action of selected anticancer drugs in follicular lymphoma B cells. Leuk. Res. *30*, 322-31.
29. Manero, F., Gautier, F., Gallenne, T., Cauquil, N., Grée, D., Cartron, P.F., Geneste, O., Grée, R., Vallette, F.M., Juin, P. (2006) The small organic compound HA14-1 prevents Bcl-2 interaction with Bax to sensitize malignant glioma cells to induction of cell death. Cancer Res. *66*, 2757-64.
30. Wang, J.L., Liu, D., Zhang, Z.J., Shan, S., Han, X., Srinivasula, S.M., Croce, C.M., Alnemri, E.S., Huang, Z. (2000) Structure-based discovery of an organic compound that binds Bcl-2 protein and induces apoptosis of tumor cells. Proc Natl Acad Sci U S A. *97*, 7124-9.

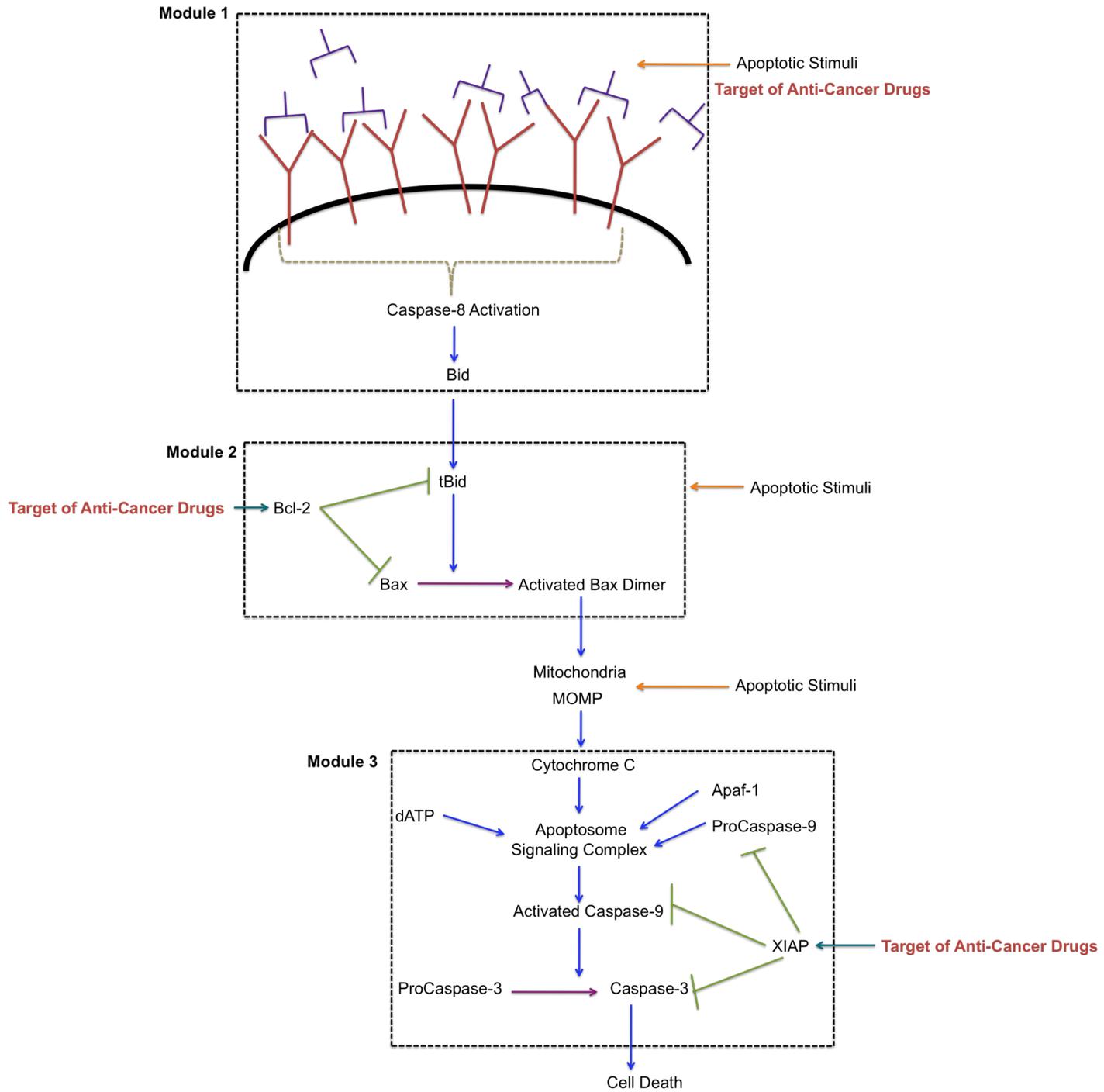

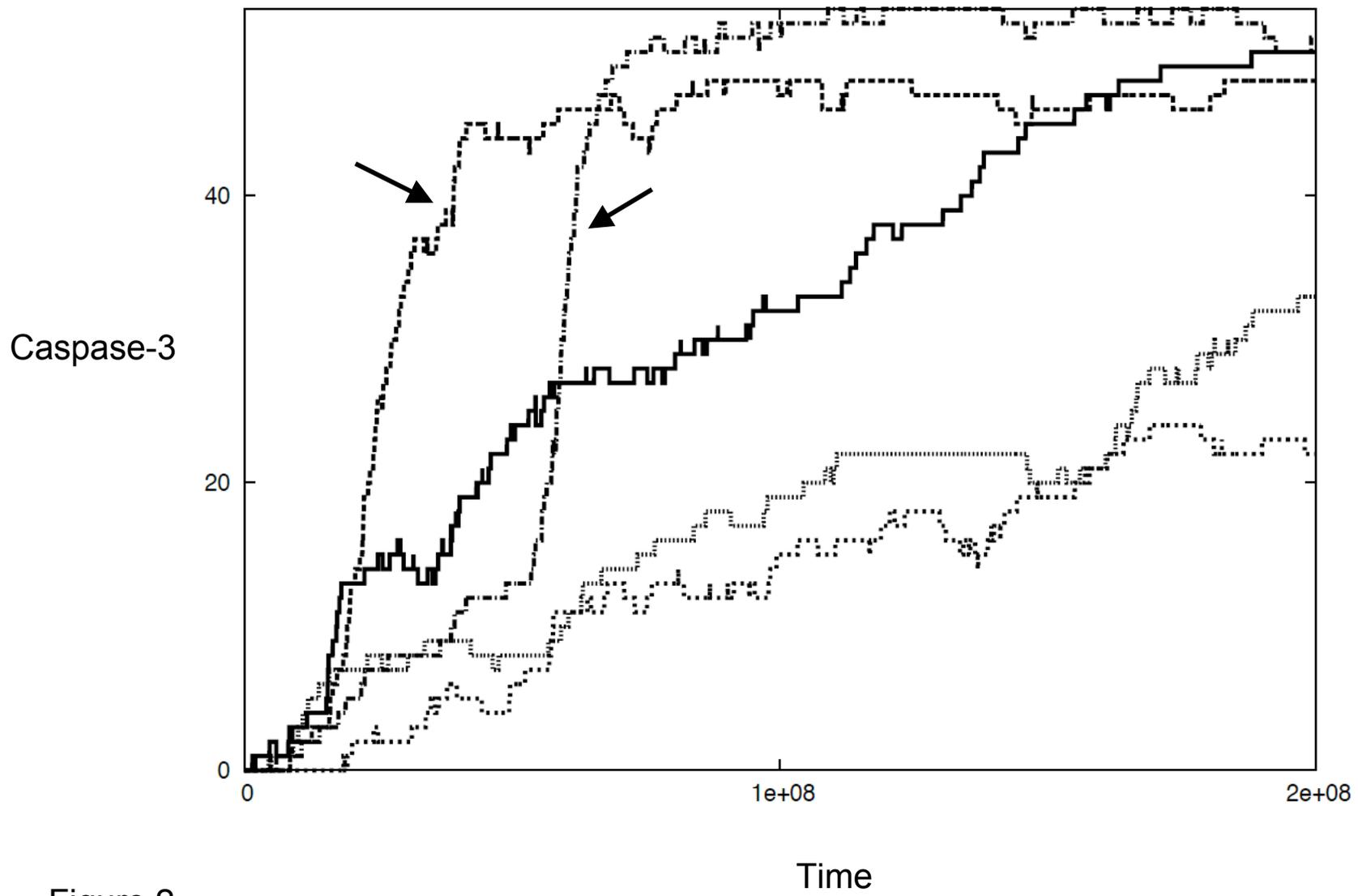

Figure 2